# Unbiased water and methanol maser surveys of NGC 1333

A-Ran Lyo, Jongsoo Kim, Do-Young Byun, and Ho-Gyu Lee
Korea Astronomy and Space Science Institute, 776, Daedeokdae-ro Yuseong-gu, Daejeon 305-348, Republic of Korea
arl@kasi.re.kr

## ABSTRACT

We present the results of unbiased 22 GHz $H_2O$ water and 44 GHz class I $CH_3OH$ methanol maser surveys in the central $7' \times 10'$ area of NGC 1333 and two additional mapping observations of a 22 GHz water maser in a $\sim 3' \times 3'$ area of the IRAS4A region. In the 22 GHz water maser survey of NGC 1333 with sensitivity of $\sigma \sim 0.3$ Jy, we confirmed masers toward $H_2O$(B) in the region of HH 7-11 and IRAS4B. We also detected new water masers at $\sim 20''$ away in the western direction of IRAS4B or $\sim 25''$ away in the southern direction of IRAS4A. We could not however find young stellar objects or molecular outflows associated with them. They showed two different velocity components of $\sim 0$ and $\sim 16$ km $s^{-1}$, which are blue- and red-shifted relative to the adopted systemic velocity of $\sim 7$ km $s^{-1}$ for NGC 1333. They also showed time variabilities in both intensity and velocity from multi-epoch observations and an anti-correlation between the intensities of the blue- and the red-shifted velocity components. We suggest that the unidentified powering source of these masers might be in the earliest evolutionary stage of star formation before the onset of molecular outflows. Finding this kind of water masers is only possible by an unbiased blind survey. In the 44 GHz methanol maser survey with sensitivity of $\sigma \sim 0.5$ Jy, we confirmed masers toward the IRAS4A2 and the eastern shock region of the IRAS2A. Both sources are also detected in 95 and 132 GHz methanol maser lines. In addition, we had new detections of methanol masers at 95 and 132 GHz toward IRAS4B. In terms of the isotropic luminosity, we detected the methanol maser sources brighter than $\sim 5\times10^{25}$ erg $s^{-1}$ from our unbiased survey.

# 1. Introduction

Outflows/jets are the most powerful products generated during the star formation process and they are the best indirect indicator of the accretion process in the circumstellar disks of low- to high-mass star formations (Arce et al. 2007 for reviews). This energetic mass-loss process has not only a constructive effect on star formation, by removing excess of angular momentum, but also a destructive effect, by changing the parent molecular cloud structures which eventually regulate star formation efficiency in the stellar clusters (Matzner & McKee 2000; Hartmann et al. 2001). Therefore, it is important to have macroscopic approaches as well as microscopic studies of the outflow to fully understand the star formation process in molecular clouds. To accomplish this, it is necessary to build up a full census of the outflows in the clouds and to interpret their activities in terms of evolutionary stages and the central stellar masses.

NGC 1333 is the most representative laboratory to study the outflow activities of low- to intermediate-mass stars thanks to their wide range of outflow evolutionary stages and its close distance of 235 pc (Hirota et al. 2008). Around 300 young stellar objects (YSOs) have been discovered, which cover the evolutionary stage from Class 0 to Class III (Lada et al. 1996; Rebull et al. 2007; Gutermuth et al. 2008). Given these advantages, many outflow survey studies have been conducted at different wavelengths, tracing different physical processes caused by outflows. These include $H_\alpha$ and [S II] optical observations as good tracers of shock-excited features in the molecular outflows (Bally et al. 1996); 2.12 μm $H_2$ near-infrared observation also as a shock tracer in the molecular outflows (Davis et al. 2008); $^{12}CO$ (J=3-2) molecular line observations as the molecular outflow tracer (Knee & Sandell 2000; Hatchell & Dunham 2009; Curtis et al. 2010); and $^{12}CO$ (J=1-0) and $^{13}CO$ (J=1-0) molecular line observations tracing molecular outflows (Arce et al. 2010; Plunkett et al. 2013). However, each tracer only provides limited information about the physical conditions of outflows. For example, the shock-excited optical $H_\alpha$ and near-infrared 2.12 μm $H_2$ features trace recently shocked gas produced by the primary jets/winds with a cooling time of only a few years, while other molecular emissions at submillimeter- to centimeter-wavelengths are produced by gas entrained and accelerated by the secondary shocks (Bally et al. 2007). Clearly, we need to obtain every detail on outflows/jets using all the available tracers to have a complete picture of them.

22 GHz $H_2O$ masers ($6_{16}$-$5_{23}$ transition) together with $H_2$ emissions are expected to arise from shocked regions excited by young high-density molecular outflows/jets, while $H_\alpha$ emissions are due to mostly atomic or ionized jets from older objects (Bally et al. 2007). Numerous studies (e.g., Haschick et al. 1980; Claussen et al. 1996) have detailed the short lifetime of $H_2O$ masers, which trace shocks and usually have lifetime measured in days for specific shock features. In particular, the VLBI (Very Large Baseline Interferometry) techniques measuring the proper motions together with the radial velocities of masers in a region as close as NGC 1333 enable us to see three dimensional motions of shocked regions. For example, $H_2O$ masers have been clearly shown for tracing the bow shocks and the inner part of the cavity walls of outflows in the massive star forming regions of AFGL 2591 and Cepheus-A from the morphologies and proper motions revealed by the VLBI multi-epoch observations (Claussen et al. 1998; Sanna et al. 2012; Torrelles et al. 2001a, 2001b, 2011, 2014; Trinidad et al. 2013). Furthermore, the 22 GHz water maser is a better tracer for the very early stage of young outflows than the near-infrared $H_2$ emission, because most early star forming regions are deeply embedded in molecular clouds.

Class I 44 GHz $CH_3OH$ methanol masers ($7_0$-$6_1$ $A^+$ transition) are also a good tracer of shocked regions caused by the interaction of the outflows with the ambient molecular clouds. 22 GHz water masers are detected from low- to high-mass YSOs (Honma et al. 2005; Moscadelli et al. 2005; Goddi & Moscadelli 2006; Moscadelli et al. 2006), while class I 44 GHz methanol masers are mainly detected from the intermediate- to high-mass YSOs (Kurtz et al. 2004; Fontani et al. 2010). Only four 44 GHz methanol masers of the low-mass star forming regions have been known until now; NGC1333-IRAS4A, NGC1333-IRAS2A, HH25MMS, and L1157 (Kalenskii et al. 2010).

We executed 22 GHz $H_2O$ water and 44 GHz $CH_3OH$ methanol maser surveys in the central 7′ ×10′ area of NGC 1333 at the center position of R.A. = $03^h$ $29^m$ $04^s$., DEC=31° 16′ 30″. (J2000) using the Korean VLBI network (KVN) 21 m telescopes. From these unbiased survey observations, we will provide a full census of water and methanol masers observed at 22 and 44 GHz frequencies within sensitivity limits of 0.3 Jy and 0.5 Jy, respectively. We also provide possible connections between the observed masers and the known outflows in the NGC 1333.

In Section 2, we describe 22 GHz $H_2O$ water maser and 44, 95, and 132 GHz class I $CH_3OH$ methanol maser observations using the KVN 21 m telescopes. In Section 3, we present the observational results. In the last section, we identify each detected maser source and briefly summarize our main results.

## 2. Observations and Data Reduction

We carried out observations of the $H_2O$ water maser line at 22.23508 GHz ($6_{16} - 5_{23}$) and class I $CH_3OH$ methanol maser lines at 44.06943 GHz ($7_0 - 6_1A^+$), 95.16946 GHz ($8_0 - 7_1A^+$), and 132.89080 GHz ($6_{-1} - 5_0E$) using the KVN 21 m telescopes over a period of 1.5 years from October 2011 to May 2013. KVN consists of three stations, at Yonsei, Ulsan, and Tamna, each of which is equipped with a multi-frequency receiver system. That enabled us to simultaneously observe four-frequency bands at 22, 43, 86, and 129 GHz with the half-power beam widths of 120″, 62″, 32″, and 23″, respectively. All observations were done with a single-dish observational mode.

In 2012, the 22 GHz water maser and 44 GHz methanol maser maps of the central 7′ × 10′ area at the center of R.A. = $03^h\ 29^m\ 04^s$. (J2000), decl. = 31° 16′ 30″. (J2000) were obtained with 60″ and 30″ grid-spacings, respectively. We carried out 22 GHz surveys in the ∼3′ × 3′ region of IRAS4A at two more epochs in 2011 and 2013 with 30″ and 15″ grid-spacings, respectively. These multi-epoch observations toward IRAS4A enabled us to increase the detectability of water masers due to their time variability. We also took single-point observations of the 22 GHz water-maser toward $H_2O$(B) source in the HH 7-11 region at multiple epochs, and the 44, 95 and 132 GHz class I methanol maser toward IRAS4A, IRAS4B and the eastern bow-shock region of IRAS2A. The pointing positions are as follows: $H_2O$(B) – R.A. = $03^h\ 29^m\ 02^s$.1., decl. = 31° 15′ 38.0″. (Haschick et al. 1980; Rodríguez et al. 1997), IRAS4A1(& 2) – R.A. = $03^h\ 29^m\ 10^s$.5. ($03^h\ 29^m\ 10^s$.4.), decl. =31◦ 13′ 31″.0″. (31◦ 13′ 32.2″.) (Looney et al. 2000), IRAS4B – R.A. = $03^h\ 29^m 12^s$.0., decl. =31° 13′ 10.0″. (Sandell et al. 1991), and the eastern shock of IRAS2A – R.A. = $03^h\ 29^m\ 00^s$.8., decl. = 31° 14′ 24.4. (Sandell & Knee 2001). All positions are at J2000 epoch.

We used a 4096-channel digital spectrometer for each frequency with a 32 MHz bandwidth. This gives the velocity coverage of 436 km $s^{-1}$ with 0.1 km $s^{-1}$ velocity resolution at 22 GHz. Pointing and focus checks have been done every 1–2 hours using the strong 22 GHz $H_2O$ and 43 GHz SiO maser lines of IK Tau (R.A. = $03^h\ 53^m\ 28^s$.87. (J2000), decl. = 11° 24′ 21.7″. (J2000)). The pointing accuracies of both bands were ≤5″ during observations for all antennas. The data were calibrated using a standard chopper wheel method and line intensities were obtained on the $T_A{}^*$ scale. The conversion factors between $T_A{}^*$ and the flux density were 12.2/12.8/13.5 Jy $K^{-1}$ for 65/62/59% (Yonsei/Ulsan/Tamna stations) aperture efficiencies at 22 GHz, 12.6/12.8/12.8 Jy $K^{-1}$ for 63/62/62% at 44 GHz, 26.5/19.9/18.5 Jy $K^{-1}$ for 39/40/43% at 95 GHz, and 33.1/38.8/24.8 Jy $K^{-1}$ for 30/32/40% at 132GHz, respectively (see the details at http://kvn.kasi.re.kr/status report/).

Table 1 shows the details of the mapping observational logs; observational dates, antenna stations, exposure integration time of each grid point, system temperatures $T_{sys}$, and root-mean-square (rms) noise levels at 0.4 km $s^{-1}$ velocity resolution.

## 3. Results

We performed the 22 GHz $H_2O$ water maser survey covering the central $7' \times 10'$ area of NGC 1333. Figure 1(a) shows a spectral map of water maser lines observed in 2012 with a 60″ grid-spacing at a center position of R.A. = $03^h$ $29^m$ $04^s.0$. (J2000), decl. = 31° 16′ 30.0″.(J2000), which is overlaid on the 2.12 μm $H_2$ image observed by Davis et al. (2008). The detected water maser lines are plotted in red color. Their positions (see Table 2) are nearby the $H_2O$(B) and IRAS4B sources, which are marked with ″+″ and ″◇″ signs in the map, respectively. Figure 1(b) and (c) show the strongest, zoom-in spectra of these detected water maser lines at each source region.

We derived the positions of detected masers using the GAUSSCLUMPS decomposition algorithm developed by Stutzki & Güsten (1990) and listed them in Table 2. The first two columns of Table 2 list maps and observed frequencies, the third and fourth columns list the positions of line intensity peaks, the fifth one lists the velocity of each line component in the frame of the local standard of rest (LSR), the sixth and seventh columns list the width and the peak flux of each line component, and the last column lists the possible maser powering source associated with each line component. Since the fitting accuracy of the peak positions determined by the GAUSSCLUMPS is far smaller than the antenna pointing accuracy, the peak position accuracy depends only on the antenna pointing accuracy, which is $\leq 5''$ for both the 22 GHz water and 44 GHz methanol maser maps. In total, three water maser line components were found at velocities of 1.05, 3.37, and 14.50 km $s^{-1}$. The derived positions of both 1.05 and 3.37 km $s^{-1}$ water maser components are in the $\sim 15''$ east of the $H_2O$(B) in the HH 7-11 region. Both of them are blue-shifted relative to the systemic velocity of 7.8 km $s^{-1}$ derived by $NH_3$ molecular line observation (Ho & Barrett 1979). The 14.50 km $s^{-1}$ water maser component comes from the nearby region of IRAS4B. It is red-shifted relative to the systemic velocity of 7.5 km $s^{-1}$ of IRAS4B (Hatchell & Dunham 2009).

We also conducted a monitoring observation toward the position of $H_2O$(B). Figure 2 shows multi-epoch 22 GHz water maser spectra of $H_2O$(B). It covers about one year from October 2011 to November 2012. The Gaussian fitting parameters of the spectra are listed in Table 3. These blue-shifted water masers show large variability in both velocity and intensity. From January 2012 to February 2012, there was a flaring event. During the interval, the peak flux of the spectral line increased from 1.43 Jy to 22.02 Jy. The highest peak flux is 131.33 Jy observed in May. After May the peak flux has decreased (see Table 3). On May 28, two strong velocity components of 1.91 km $s^{-1}$ and 3.33 km $s^{-1}$ appeared with peak fluxes of 131.33 Jy and 86.55 Jy, respectively. In June, the more blue-shifted component is accelerated further, while the less blue-shifted one is decelerated relative to the systemic velocity, 7.8 km

s$^{-1}$. If we looked at the LSR velocities at peaks of line profiles over the whole observation interval, the LSR velocity is 2.36 km s$^{-1}$ on October 11, 2011, keeps increasing (deceleration relative to the systemic velocity) up to 3.82 km s$^{-1}$ on March 24, 2012, and then decreases (acceleration relative to the systemic velocity) to 2.82 km s$^{-1}$ at the end of our observations. It seems that the change from deceleration to acceleration is related to the occurrence of multiple strong masers on May 28, 2012.

In 2011 before the main surveys, we mapped a small $2' \times 3'$ area of IRAS4A with a $30''$ grid-spacing. Figure 3(a) and (b) show a water maser spectral map and a zoom-in spectrum, respectively. One maser component was found at the velocity of 7.04 km s$^{-1}$ close to IRAS4A at the south-eastern part, with a separation of $\sim 10''$. It is slightly blue-shifted compared to the systemic velocity of 7.6 km s$^{-1}$ (Hatchell & Dunham 2009). Since the derived position of the 7.04 km s$^{-1}$ component is coincident with those of water masers observed using the VLBA, its driving source might be the IRAS4A2. However, the distribution of water maser features is nearly perpendicular to the CO outflow direction, it still needs further investigation on the origin of the component (Marvel et al. 2008).

In 2013, we observed again the $3' \times 3'$ area of IRAS4A with a $20''$ grid-spacing and found new water maser components with velocities at -0.01 km s$^{-1}$ and 16.32 km s$^{-1}$ (Fig. 3(c)). These newly discovered components are located in the western part of the IRAS4B with $\sim 20''$ separation or $\sim 25''$ away in the southern direction of IRAS4A based on their positions by using the GAUSSCLUMPS (see Table 2). We also carried out multi-epoch observations on this source during less than the two-month period from April to May in 2013. Figure 3(d) showed variabilities in both velocity and intensity. It also showed an anti-correlation between blue- and red-shifted velocity components in their intensities. In other words, the blue-shifted maser is getting weaker while the red-shifted one becomes stronger.

In Figure 4, we showed the 44 GHz $CH_3OH$ methanol maser map with a $30''$ grid-spacing covering the same area as the water maser map of Figure 1(a). Again the detected line profiles at the nearby regions of IRAS4A1, IRAS4A2 (marked with "×" and "△" signs) and IRAS2A (marked with "□" sign) are plotted with red color. We found two maser components at velocities of 6.65 and 9.37 km s$^{-1}$, and their positions derived using the GAUSSCLUMPS are listed in Table 2. The slightly blue-shifted 6.65 km s$^{-1}$ velocity component with respect to the systemic velocity is from $\sim 14''$ separation in the south-western direction of IRAS4A. The direction is consistent with that of the blue-shifted CO (J=3-2) outflow of IRAS4A2 (Knee & Sandell 2000). The maser emission of 9.37 km s$^{-1}$ comes from the eastern bow-shock region of IRAS2A at the separation of $\sim 15''$ eastern direction. It is red-shifted relative to the systemic velocity 7.7 km $s^{-1}$ of IRAS2A (Hatchell & Dunham 2009).

We confirmed that all 44, 95, and 132 GHz methanol masers are from the nearby region of IRAS4A at $\sim$10–19$''$ separation in the south-western direction based on the mappings of a 2$'$ × 2$'$ IRAS4A area with 15$''$ grid-spacing (we do not present all these maps here). The positions of all these maser components derived using GAUSSCLUMPS are consistent with that of the blue-shifted south-western outflow of IRAS4A2. The representative spectra of all bands are shown in the Figure 5(a). We found two Gaussian components for each frequency; one 4.52 km $s^{-1}$ velocity component with a broad line width δv $\sim$9.42 km $s^{-1}$ and another 6.97 km $s^{-1}$ component with a narrow line width δv $\sim$1.54 km $s^{-1}$ at 44 GHz; a broad line of 5.28 km $s^{-1}$ with δv $\sim$7.90 km $s^{-1}$ and a narrow line of 7.08 km $s^{-1}$ with δv $\sim$0.62 km $s^{-1}$ at 95 GHz; a broad line of 3.98 km $s^{-1}$ with δv $\sim$9.38 km $s^{-1}$ and a narrow line of 6.69 km $s^{-1}$ with δv $\sim$2.32 km $s^{-1}$ at 132 GHz band. The line parameters of all the Gaussian components of IRAS4A2 are listed in Table 4. The detection of a 132 GHz class I methanol maser is the first time on this source, while 44 and 95 GHz masers were detected by Kalenskii et al. (2010). We also clearly found broad wing features of δv $\geq$8 km $s^{-1}$ at all frequencies, which might be thermal components due to the molecular outflow itself.

With position-switching mode observations, we also detected all 44, 95, and 132 GHz methanol maser lines toward IRAS4B, which are shown in the Figure 5(b). Since the separation angle of $\sim$ 30$''$ between IRAS4B and IRAS4A is larger than the half-beam sizes of $\sim$ 16$''$ at 95 GHz and $\sim$ 12$''$ at 132 GHz, respectively, we resolved each region at 95 and 132 GHz observations. In fact, the 95 and 132 GHz methanol maser line profiles from IRAS4B are different from those of IRAS4A2. We detected for the first time class I methanol masers at 95 and 132 GHz toward IRAS4B. However, we could not resolve each region at 44 GHz observation since the half-beam size of 44 GHz observation is similar to the angular separation. The 44 GHz line profiles of IRAS4B and IRAS4A2 shown in Figure 5(a) and (b) are similar and there is very little difference in their LSR velocities listed in Table 4. We again found two Gaussian components toward IRAS4B for each frequency; a broad line of 4.84 km $s^{-1}$ with δv $\sim$10.39 km $s^{-1}$ and a narrow line of 6.67 km $s^{-1}$ with δv $\sim$1.82 km $s^{-1}$ at 44 GHz; a broad line of 6.66 km $s^{-1}$ with δv $\sim$24.57 km $s^{-1}$ and a narrow line of 7.17 km $s^{-1}$ with δv $\sim$3.06 km $s^{-1}$ at 95 GHz; a broad line of 5.45 km $s^{-1}$ with δv $\sim$14.90 km $s^{-1}$ and a narrow line of 7.69 km $s^{-1}$ with δv $\sim$3.16 km $s^{-1}$ at 132 GHz band. The line parameters of all the Gaussian components of IRAS4B are listed in Table 4. We suggest that all these broad wing features at all frequencies might also be due to the molecular outflow itself

Figure 5(c) show all methanol maser lines toward the eastern shocked region of IRAS2A, whose central velocities around 10 km $s^{-1}$ (see Table 4) are consistent with those in Kalenskii et al. (2010). We found two Gaussian components for each frequency; a 10.03 km $s^{-1}$ with $\delta v$ ~1.21 km $s^{-1}$ and a 10.03 km $s^{-1}$ with $\delta v$ ~0.81 km $s^{-1}$ at 44 GHz, a3.85kms$^{-1}$ with$\delta v$~3.36kms$^{-1}$ anda2.94kms$^{-1}$ with $\delta v$~1.29kms$^{-1}$ at 95GHz, and two broad components of 2.87 and 5.14 km $s^{-1}$ with $\delta v$ >9 km $s^{-1}$ at 132 GHz band. The line parameters of the Gaussian components of IRAS2A are listed in Table 4. We also newly detected a 132 GHz methanol maser, which is much stronger than those of 44 and 95 GHz. For example, the integrated maser luminosity of 132 GHz is almost 10 times stronger than that of 95 GHz, while IRAS4A2 and IRAS4B showed only 3 times stronger at 132 GHz than at 95 GHz.

## 4. Summary and Discussions

Detections of 22 GHz water masers and 44 GHz class I methanol masers are good indicators of the current ongoing star formation process because they are closely related with outflows/jets activities (Furuya et al. 2003; Kalenskii et al. 2006, 2010). In particular, the short time variabilities of the water masers in both intensity and velocity support the idea that observations of masers are the best way to watch the vivid live-show of outflow activities.

In 2012, we carried out the unbiased 22 GHz $H_2O$ water and 44 GHz class I $CH_3OH$ methanol maser surveys in the central $7' \times 10'$ area of NGC 1333 in 2012 using the KVN telescopes with sensitivities of $\sigma$ ~0.3 Jy at the 22 GHz map and $\sigma$ ~0.5 Jy at the 44 GHz map (see Fig. 1(a) and Fig. 4). In addition, we carried out mapping observations of a 22 GHz water maser in the area of IRAS4A at two more epochs in 2011 and 2013 to see the variability in velocity and intensity. We also made 44, 95, and 132 GHz methanol maser observations toward IRAS4A, IRAS4B and IRAS2A regions with the position-switching mode of observations.

We summarize our results as follows. Firstly, in the 22 GHz water maser survey done in 2012, we confirmed the well-known 22 GHz water masers of $H_2O$(B) in the HH 7-11 region shown in Figure 1 (Haschick et al. 1980; Claussen et al. 1996; Rodríguez et al. 1997; Wootten et al. 2002). The monitoring observations of the 22 GHz water masers of $H_2O$(B) show high variability in both intensity and velocity. The LSR velocities of the peak of water maser line profiles initially decelerate with respect to the systemic velocity, and then accelerate when multiple strong masers occur (Fig. 2). In particular, during the 10 days' interval of two epochs, May and June 2012, the more blue-shifted component was accelerated further, while the less blue-shifted one was decelerated relative to the systemic velocity. The line parameters of 10 epoch observations are listed in Table 3. We also confirmed the water maser of IRAS4B with a velocity of 14.50 km $s^{-1}$ during the 2012 survey observation (Fig. 1(c)). Secondly, in the 2011 survey map shown in Figure 3(a), we confirmed the slightly blue-shifted 7.04 km $s^{-1}$ water maser of IRAS4A2, which firstly reported by Claussen et al. (1996). The derived position measured using GAUSSCLUMPS is consistent with those of water masers observed by Marvel et al. (2008). However, the distribution of water maser features is nearly perpendicular to the CO outflow direction, it still needs further investigation on the origin of the component.

Thirdly, in the 2013 unbiased spectral survey shown in Figure 3(c), we discovered two new water masers at positions $\sim 20'''$ away in the western direction of IRAS4B ($\sim 25''$ away in the southern direction of IRAS4A) (see also Table 2 of the estimated positions measured using GAUSSCLUMPS). From multi-epoch observations, the new water masers show time variabilities in both intensity and velocity. In addition, it seems that there is an anti-correlation in their intensities between blue- and red-shifted velocity components (Fig. 3(d)). This indicates that both velocity components are genuinely related to each other. However, we could not find any clear candidate for the powering source for these water masers from sub(millimter) continuum observations and no sign of outflows/jets from the previous molecular line survey studies. We found only one possible continuum candidate with only 5σ level at 96.7 GHz band observed by Sakai et al. (2012; see Fig. 1 in their paper), which is around 5–10$''$ away in the north direction from these maser components. We need further deep (sub)millimeter observations to confirm whether this weak continuum emission is real or not. It will provide information whether it is indeed a deeply embedded earliest stage of YSO before the onset of other molecular outflows.

Fourthly, in the 44 GHz methanol maser survey shown in Figure 4, we confirmed the known maser sources of the IRAS4A and the eastern shock region of IRAS2A (Kalenskii et al. 2006, 2010). We also detected 95 and 132 GHz methanol masers toward both sources (Fig. 5). In particular, we found that the derived positions of all blue-shifted methanol masers with velocity ranges of $\sim$3–7 km $s^{-1}$ at 44, 95, and 132 GHz in IRAS4A region are consistent with that of the south-western blue-shifted outflow of IRAS4A2. That is the first detection of 132 GHz class I methanol masers of both objects from our observation. We found that the newly detected 132 GHz methanol maser luminosity on IRAS2A is almost 10 times stronger than that of 95 GHz, while IRAS4A2 showed only 3 times stronger at 132 GHz than 95 GHz in its luminosity. In both sources, the broad wing features shown in the all frequencies might be the thermal emissions from the molecular outflow itself.

Lastly, we have a possible new detection of methanol masers of IRAS4B by the position-switching observational mode at 44, 95 and 132 GHz (see Fig. 5(b) and Table 4 for the line fitting parameters). We suggested that 95 and 132 GHz methanol maser emissions originated from the IRAS4B based on their different line profiles compared to those of IRAS4A2 in spite of their close distance of $\sim 30''$; the half-beam sizes of these bands are less than their separation angle.

We detected, through our 44 GHz methanol maser survey toward the central region of NGC 1333, maser sources brighter than $\sim 5 \times 10^{25}$ erg $s^{-1}$ of the isotropic maser luminosity. The 1σ sensitivity of our 44 GHz mapping observations is about 0.5 Jy. The peak fluxes of detected masers are above around 2 Jy, which is about the 4σ level. We listed the integrated flux over the velocity of each Gaussian component in Table 4. The isotropic luminosity of each component was calculated by adopting the distance of 235 pc to NGC 1333. The minimum isotropic luminosity of the detected methanol masers in our survey is $2.86 \times 10^{-8} L_{\odot}$, which is from the eastern shock region of IRAS2A (see also Table 4).

We are grateful to all staff members at the KVN who helped to operate. The KVN is a facility operated by the Korea Astronomy and Space Science Institute.

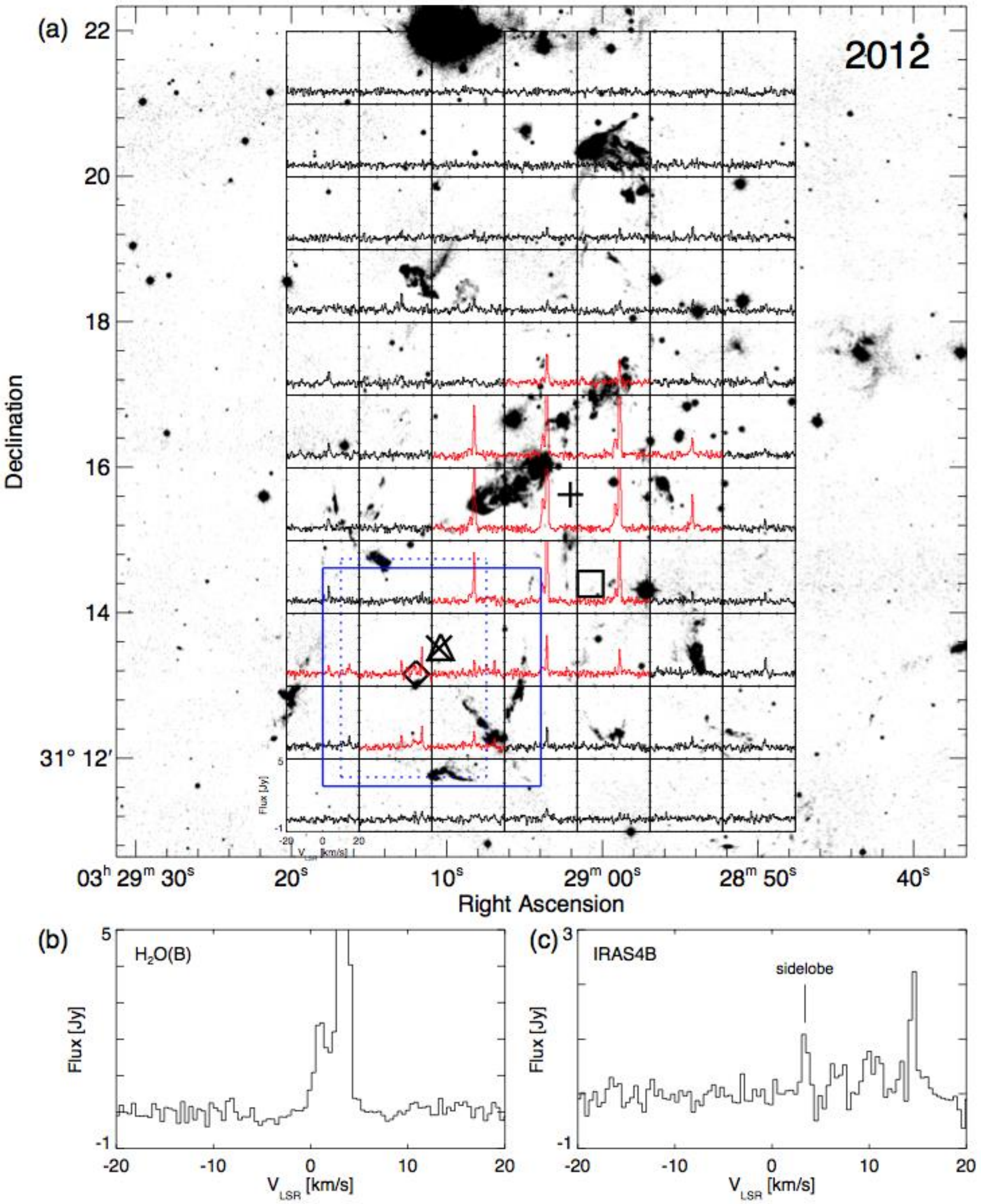

Figure 1 - (a) 22 GHz $H_2O$ water maser map observed in 2012 covers the central $7' \times 10'$ area of the NGC 1333 with a 60" grid-spacing at the center position of R.A. = 03 29 04. (J2000), decl. = 31 16 30. (J2000), which is overlaid on 2.12 μm $H_2$ image observed by Davis et al. (2008). Spectra are smoothed to have a 0.4 km $s^{-1}$ velocity resolution and the rms noise level is 0.014 K. Symbols of ″+″, ″×″, ″△″, ″◇″, and ″□″ mark positions of $H_2O$(B), IRAS4A1, IRAS4A2, IRAS4B, and the eastern shock region of IRAS2A, respectively. The detected spectra are highlighted with red-color, except the side-lobe spectra due to the strong emission from $H_2O$(B). A label for each spectrum is marked in the left-bottom panel. x-axis is the velocity in units of km $s^{-1}$ and y-axis is the flux in units of Jy. The dotted- and solid-line boxes with blue-color represent survey areas of water maser done in two different epochs on 2011 (Fig. 3a) and on 2013 (Fig. 3c), respectively. (b) The enlarged spectrum toward the $H_2O$(B)(″+″). (c) The enlarged spectrum toward the IRAS4B(″3″).

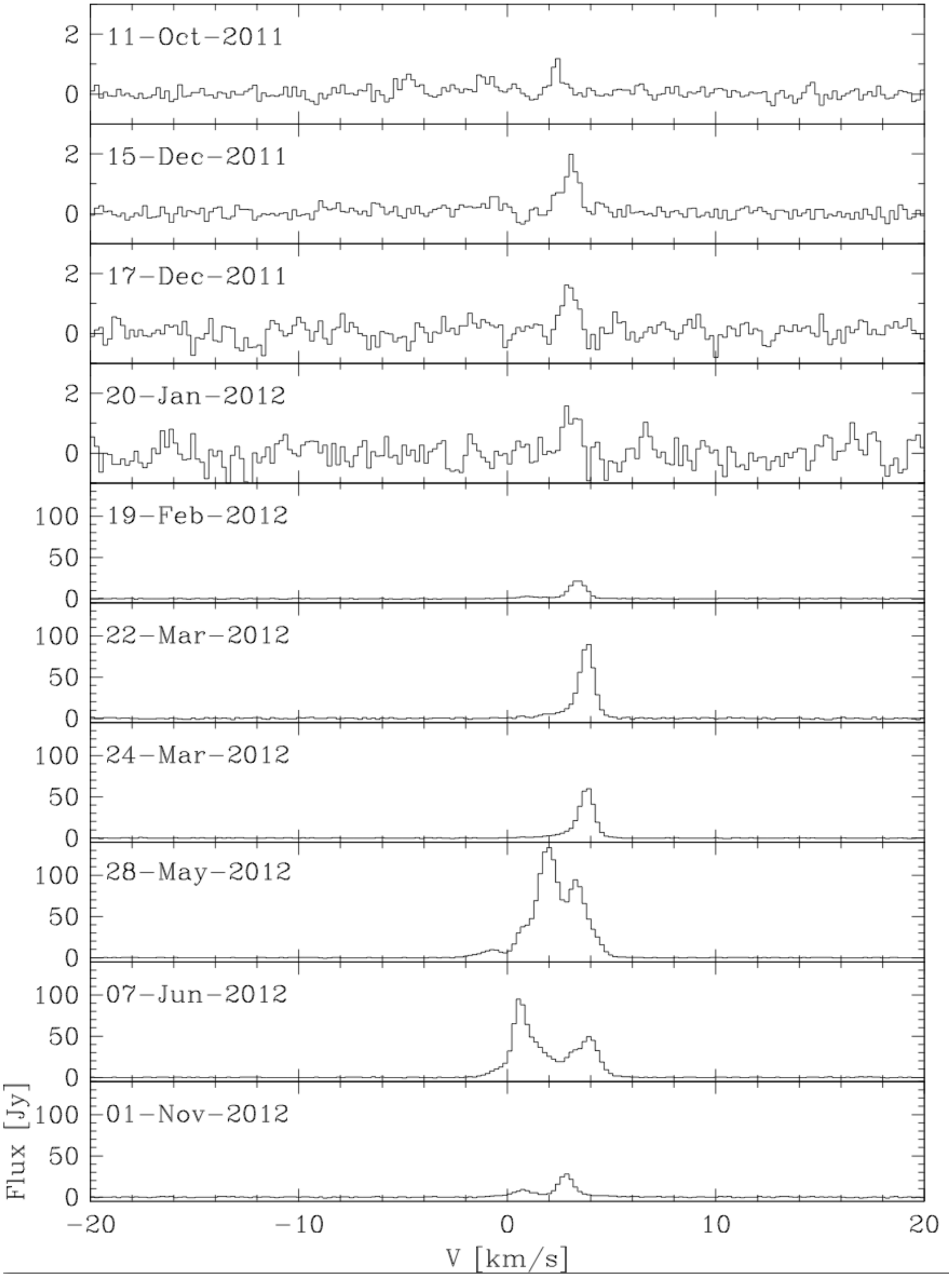

Figure 2 - Multi-epoch 22 GHz water maser spectra of $H_2O$(B) in the HH 7-11 region. Please note that the vertical scale has been changed from January 20, 2012 to February 19, 2012**.**

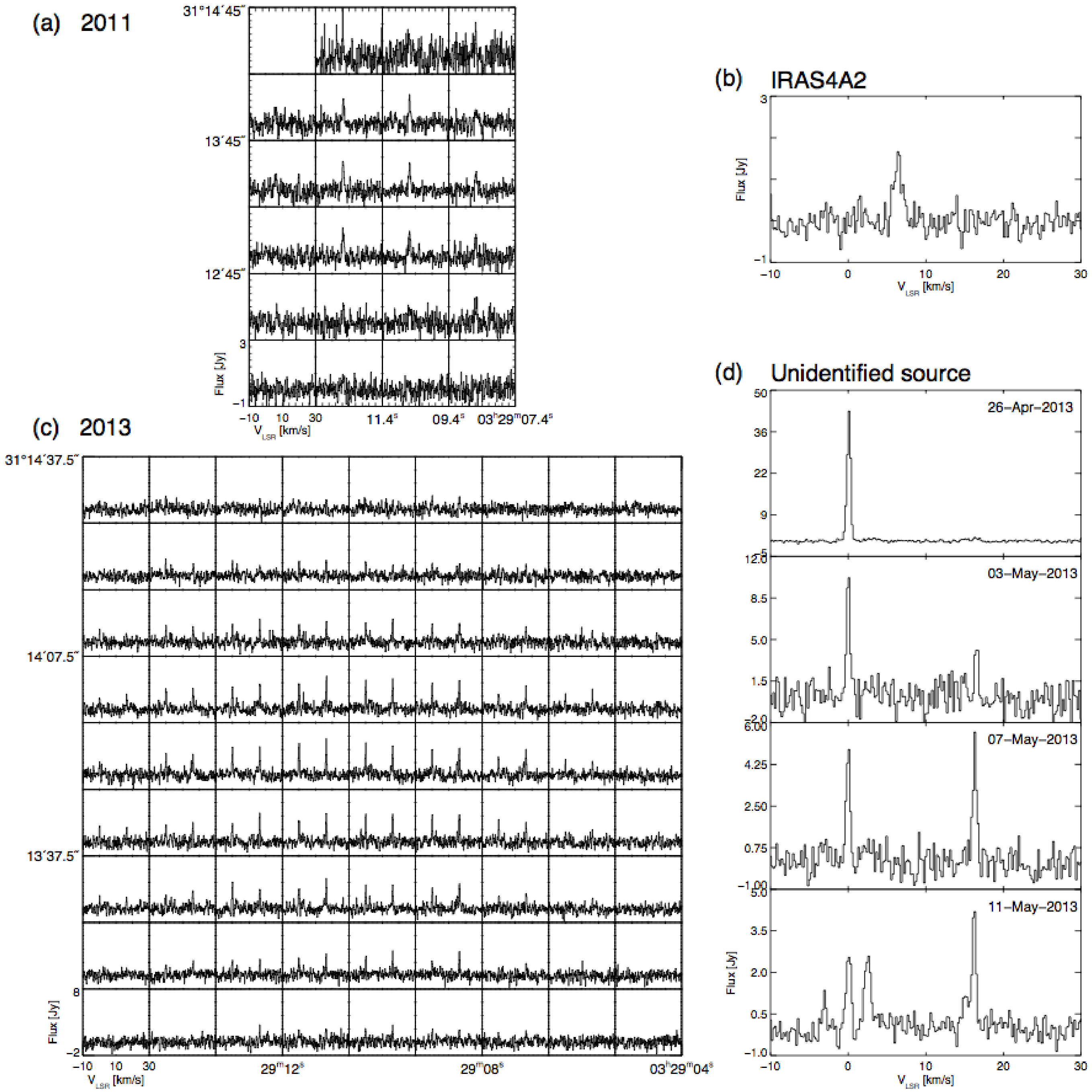

Figure 3 - (a) 22 GHz H2O water maser map observed in 2011 covers 2′ × 3′ area (blue dotted-line box in Fig. 1a) of the IRAS4A region with a 30″ grid-spacing. The rms noise level is 0.03 K at 0.2 km $s^{-1}$ velocity resolution. (b) The enlarged spectrum shown in the map (a), which is identified as IRAS4A2. (c) 22 GHz H2O water maser map observed in 2013 covers 3′ × 3′ area (blue solid-line box in Fig. 1a) of the IRAS4A region with a 20″ grid-spacing. The rms noise level is 0.04 K at 0.2 km $s^{-1}$ velocity resolution. (d) The multi-epoch spectra of the maser shown in the map (c).

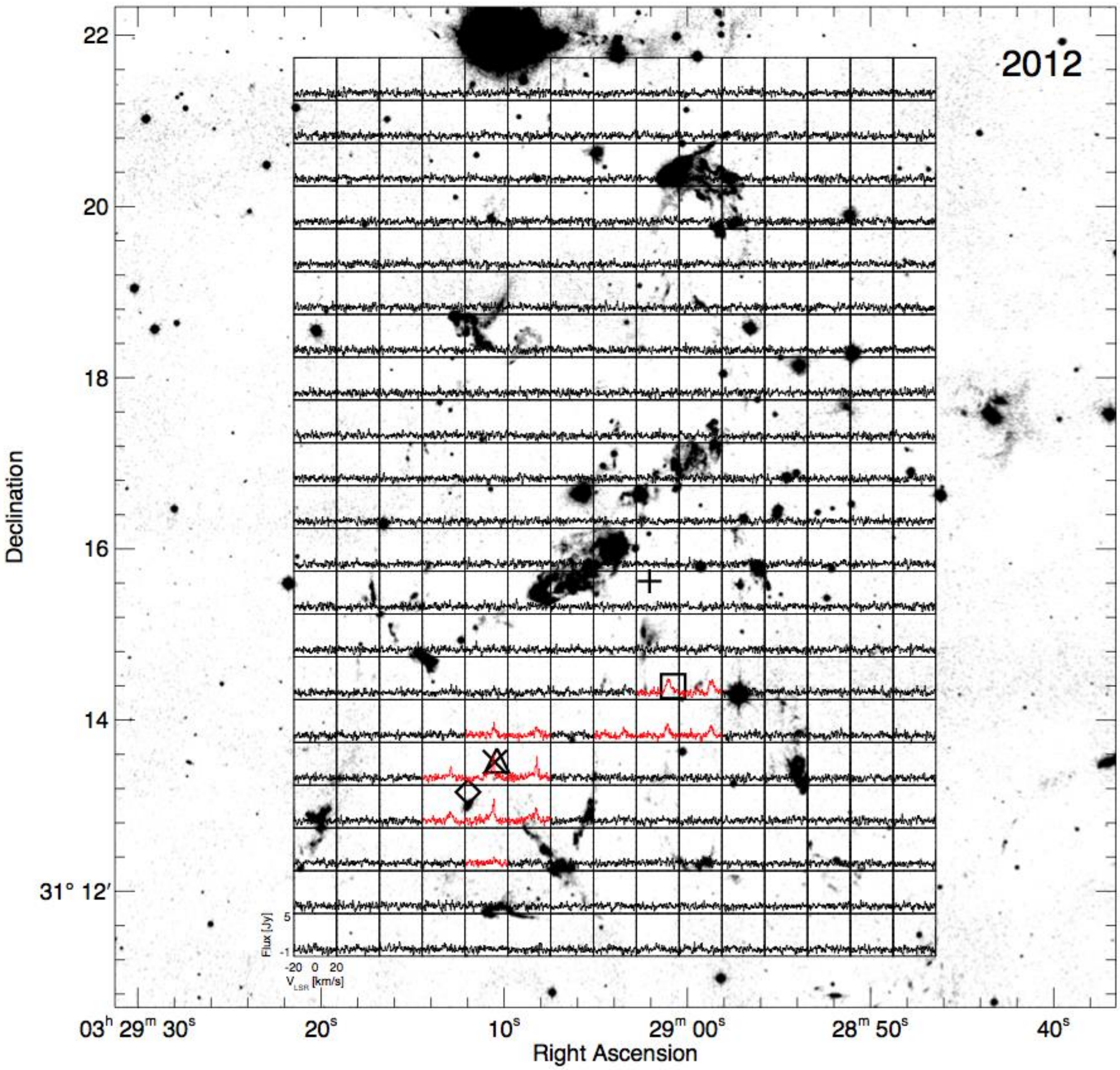

Figure 4 - 44 GHz $CH_3OH$ methanol maser map observed in 2012 covers the same area as Fig. 1(a) and it is also overlaid on 2.12 μm $H_2$ image observed by Davis et al. (2008). All the labels and signs are same as Fig. 1(a). The rms noise level is 0.028 K at 0.4 km $s^{-1}$ velocity resolution. The line-detected spectra are highlighted with red-color in the IRAS4A region and the eastern shock region of the IRAS2A.

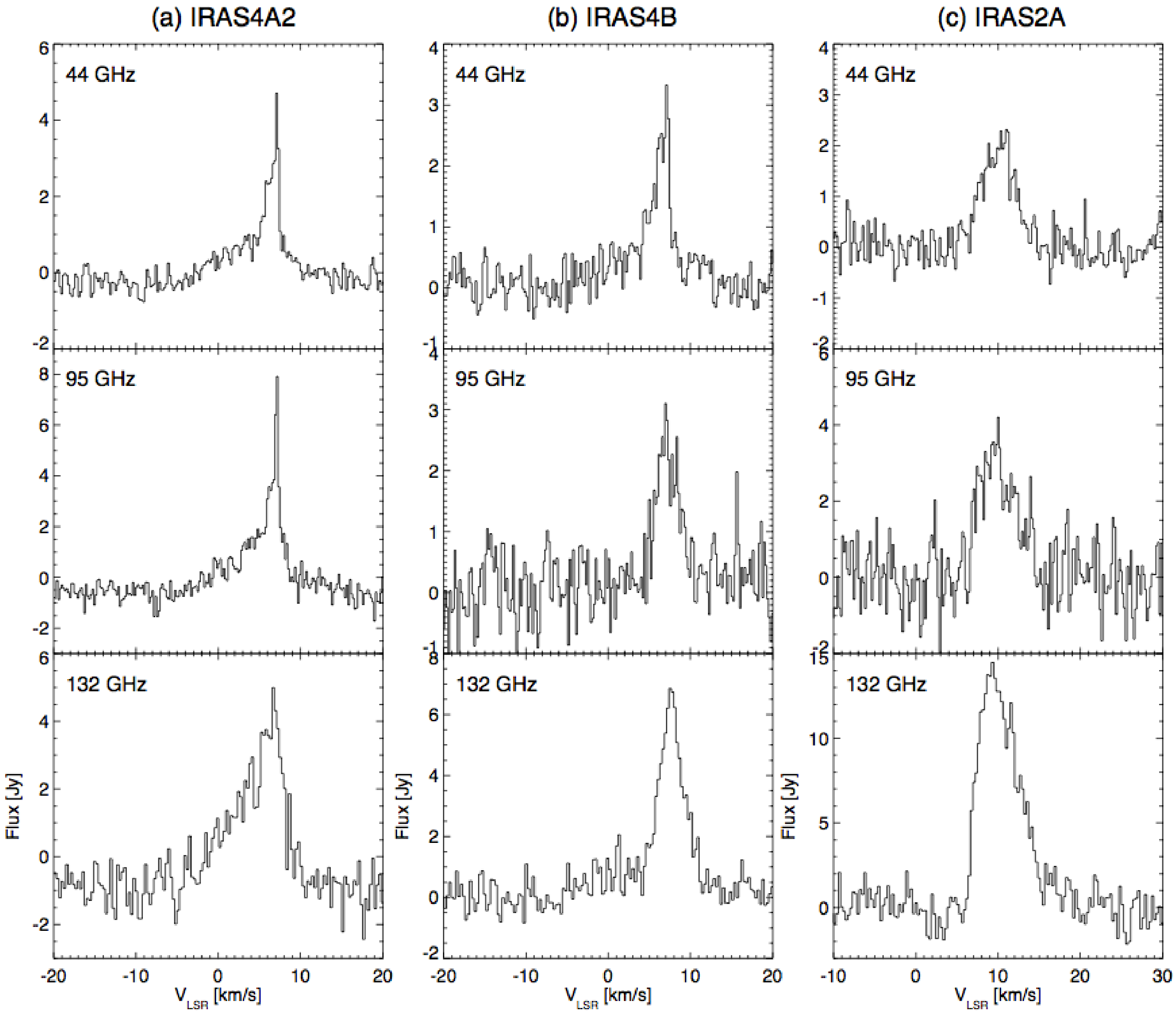

Figure 5 - (a) IRAS4A2 spectra of 44, 95, and 132 GHz methanol masers. (b) IRAS4B spectra of 44, 95, and 132 GHz methanol masers. (c) The spectra of the IRAS2A eastern shock region of 44, 95, and 132 GHz methanol masers.

Table 1: Observational logs

| Map | Fig 1 | Fig 3a | Fig 3b | Fig 4 |
|---|---|---|---|---|
| frequency (GHz) | 22.23508 | 22.23508 | 22.23508 | 44.06943 |
| Date of observation | 2012 Feb 19-21 | 2011 Oct 13 | 2013 May 07 | 2012 Jan 07-11 |
| Antenna station | Ulsan | Yonsei | Yonsei | Ulsan |
| Integration time (minutes) | 42 | 30 | 10 | 19 |
| $T_{sys}$ (K) | 73 | 120 | 100 | 140 |
| rms noise level$^{\star}$ (K) | 0.01 | 0.02 | 0.03 | 0.03 |

$\star$ at 0.4 km s$^{-1}$ velocity resolution

Table 2: Fitting parameters from GAUSSCLUMPS

| Map | Frequency (GHz) | RA (J2000) | Dec. (J2000) | $V_{LSR}$ (km s$^{-1}$) | $\delta v$ (km s$^{-1}$) | $F_{peak}$ (Jy) | possible identified source |
|---|---|---|---|---|---|---|---|
| Fig 1 | 22.23508 | 03 29 02.80 | 31 15 38.4 | 1.05 | 1.38 | 3.0 | $H_2O$(B) (HH 7-11 region) |
| | | 03 29 02.96 | 31 15 33.6 | 3.37 | 0.95 | 21.2 | $H_2O$(B) (HH 7-11 region) |
| | | 03 29 11.72 | 31 13 07.8 | 14.50 | 0.49 | 3.2 | IRAS4B |
| Fig 3a | 22.23508 | 03 29 10.92 | 31 13 23.4 | 7.04 | 1.08 | 1.4 | IRAS4A2 |
| Fig 3b | 22.23508 | 03 29 10.48 | 31 13 10.8 | -0.01 | 0.63 | 5.0 | unidentified source |
| | | 03 29 10.84 | 31 13 06.6 | 16.32 | 0.64 | 5.0 | unidentified source |
| Fig 4 | 44.06943 | 03 29 09.68 | 31 13 21.6 | 6.65 | 1.90 | 4.0 | IRAS4A2 |
| | | 03 29 01.40 | 31 14 18.6 | 9.37 | 4.59 | 2.2 | IRAS2A eastern shock |

Table 3: Gaussian fitting parameters of $H_2O$(B) water maser lines shown in Figure 2

| Date (JD) | $\int F_\nu dv$ (Jy km $s^{-1}$) | $v_{LSR}$ (km $s^{-1}$) | $\delta v$ (km $s^{-1}$) | $F_{peak}$ (Jy) |
|---|---|---|---|---|
| 12-Oct-2011(2455847.45065) | 0.66±0.10 | 2.36±0.04 | 0.55±0.12 | 1.13 |
| 15-Dec-2011(2455911.27931) | 1.68±0.14 | 3.05±0.04 | 0.94±0.11 | 1.68 |
| 17-Dec-2011(2455913.27969) | 1.43±0.18 | 2.97±0.06 | 0.81±0.11 | 1.66 |
| 20-Jan-2012(2455947.12510) | 1.28±0.20 | 3.03±0.07 | 0.84±0.13 | 1.43 |
| 19-Feb-2012(2455977.07182) | 4.71±0.23 | 1.18±0.04 | 1.79±0.10 | 2.47 |
| | 21.07±0.17 | 3.37±0.00 | 0.90±0.01 | 22.02 |
| 22-Mar-2012(2456008.84657) | 25.53±1.04 | 3.10±0.05 | 2.28±0.07 | 10.51 |
| | 66.66±0.88 | 3.82±0.00 | 0.76±0.01 | 82.64 |
| 24-Mar-2012(2456010.87586) | 17.64±0.74 | 3.28±0.04 | 1.93±0.06 | 8.59 |
| | 43.06±0.70 | 3.82±0.00 | 0.75±0.01 | 54.32 |
| 28-May-2012(2456075.64072) | 12.74±0.24 | -0.80±0.01 | 1.32±0.03 | 9.07 |
| | 30.99±0.30 | 0.76±0.00 | 0.85±0.01 | 34.07 |
| | 140.16±0.42 | 1.91±0.00 | 1.00±0.00 | 131.33 |
| | 126.05±0.39 | 3.33±0.00 | 1.37±0.01 | 86.55 |
| 07-Jun-2012(2456085.69998) | 40.01±0.36 | 0.59±0.00 | 0.61±0.00 | 61.76 |
| | 98.15±0.42 | 1.14±0.01 | 2.24±0.01 | 41.10 |
| | 67.39±0.32 | 3.80±0.00 | 1.38±0.01 | 45.96 |
| 01-Nov-2012(2456233.10977) | 1.05±0.35 | -0.49±0.13 | 0.82±0.23 | 1.20 |
| | 7.64±0.78 | 0.72±0.02 | 1.00±0.08 | 7.16 |
| | 20.23±0.54 | 2.82±0.00 | 0.78±0.02 | 24.47 |
| | 13.18±1.23 | 2.83±0.16 | 3.75±0.33 | 3.31 |

Table 4: Gaussian fitting parameters of detected lines shown in Figure 3 and Figure 5

| Source | Frequency (GHz) | Date of observation | $\int F_\nu dv$ (Jy km s$^{-1}$) | $v_{\rm LSR}$ (km s$^{-1}$) | $\delta v$ (km s$^{-1}$) | $F_{\rm peak}$ (Jy) | $L^{a}_{\rm comp}$ ($L_\odot$) | $L^{b}_{\rm sum}$ ($L_\odot$) |
|---|---|---|---|---|---|---|---|---|
| unidentified | 22.23508 | 26-Apr-2013 | 26.40±0.20 | 0.09±0.00 | 0.56±0.01 | 44.64 | 3.92E-8 | 3.92E-8 |
| source | 22.23508 | 03-May-2013 | 6.50±0.60 | 0.01±0.02 | 0.54±0.06 | 11.24 | 9.65E-9 | |
| | | | 2.95±0.54 | 16.55±0.05 | 0.55±0.10 | 5.07 | 4.38E-9 | 1.40E-8 |
| | 22.23508 | 07-May-2013 | 3.33±0.28 | -0.01±0.03 | 0.63±0.06 | 4.96 | 4.95E-9 | |
| | | | 3.41±0.29 | 16.32±0.03 | 0.64±0.07 | 5.02 | 5.06E-9 | 1.00E-8 |
| | 22.23508 | 11-May-2013 | 0.61±0.13 | -3.05±0.05 | 0.42±0.10 | 1.37 | 9.06E-10 | |
| | | | 2.13±0.17 | 0.07±0.03 | 0.72±0.06 | 2.79 | 3.16E-9 | |
| | | | 2.66±0.21 | 2.51±0.04 | 0.95±0.09 | 2.62 | 3.95E-9 | |
| | | | 1.20±0.23 | 15.16±0.10 | 0.95±0.22 | 1.19 | 1.78E-9 | |
| | | | 2.63±0.20 | 16.24±0.02 | 0.55±0.05 | 4.48 | 3.91E-9 | 1.37E-8 |
| IRAS4A1 | 22.23508 | 13-Oct-2011 | 1.61±0.21 | 7.04±0.05 | 1.08±0.21 | 1.40 | 2.39E-9 | 2.39E-9 |
| IRAS4A2 | 44.06943 | 11-Mar-2012 | 10.89±0.63 | 4.52±0.29 | 9.42±0.58 | 1.09 | 3.08E-8 | |
| | | | 4.41±0.35 | 6.79±0.04 | 1.54±0.12 | 2.69 | 1.25E-8 | 4.33E-8 |
| | 95.16946 | 11-Mar-2012 | 20.95±0.82 | 5.28±0.16 | 7.90±0.49 | 2.49 | 2.08E-7 | |
| | | | 3.80±0.46 | 7.08±0.02 | 0.62±0.11 | 5.73 | 3.78E-8 | 2.46E-7 |
| | 132.89080 | 11-Mar-2012 | 29.68±2.02 | 3.98±0.33 | 9.38±0.63 | 2.97 | 5.12E-7 | |
| | | | 7.85±1.44 | 6.69±0.08 | 2.32±0.33 | 3.18 | 1.35E-7 | 6.47E-7 |
| IRAS4B | 22.23508 | 20-Feb-2012 | 0.64±0.20 | 9.83±0.08 | 0.52±0.20 | 1.17 | 9.51E-10 | |
| | | | 0.65±0.20 | 10.96±0.13 | 0.69±0.24 | 0.89 | 9.65E-10 | |
| | | | 0.37±0.17 | 13.29±0.07 | 0.41±0.19 | 0.84 | 5.50E-10 | |
| | | | 1.65±0.18 | 14.51±0.03 | 0.49±0.06 | 3.18 | 2.45E-9 | 4.92E-9 |
| | *44.06943*[c] | *26-Apr-2012* | *7.79±0.72* | *4.84±0.46* | *10.39±1.07* | *0.70* | *2.20E-8* | |
| | | | *3.96±0.44* | *6.67±0.07* | *1.82±0.18* | *2.04* | *1.12E-8* | *3.32E-8* |
| | 95.16946 | 26-Apr-2012 | 7.36±2.17 | 6.66±0.59 | 24.57±10.69 | 0.28 | 7.32E-8 | |
| | | | 7.23±0.86 | 7.17±0.13 | 3.06±0.32 | 2.22 | 7.19E-8 | 1.45E-7 |
| | 132.89080 | 26-Apr-2012 | 15.00±1.55 | 5.45±0.71 | 14.90±1.37 | 0.95 | 2.59E-7 | |
| | | | 18.29±1.05 | 7.69±0.05 | 3.16±0.15 | 5.44 | 3.15E-7 | 5.74E-7 |
| eastern shocked | 44.06943 | 20-Mar-2012 | 6.02±0.42 | 10.03±0.17 | 4.66±0.27 | 1.21 | 1.70E-8 | |
| region of IRAS2A | | | 4.11±0.15 | 10.03±0.27 | 4.77±0.62 | 0.81 | 1.16E-8 | 2.86E-8 |
| | 95.16946 | 20-Mar-2012 | 13.75±3.34 | 9.25±0.44 | 3.85±0.78 | 3.36 | 1.37E-7 | |
| | | | 4.03±2.97 | 12.83±0.96 | 2.94±1.20 | 1.29 | 4.01E-8 | 1.77E-7 |
| | 132.89080 | 20-Mar-2012 | 27.94±5.74 | 8.56±0.10 | 2.87±0.22 | 9.14 | 4.82E-7 | |
| | | | 52.16±6.29 | 11.24±0.29 | 5.14±0.36 | 9.54 | 9.00E-7 | 1.38E-6 |

[a] isotropic maser luminosity of each velocity component estimated by adopting the distance of 235 pc to NGC 1333

[b] Summed luminosity of all the velocity components

[c] It could have contamination from IRAS4A.

# REFERENCES

Arce, H. G., Borkin, M. A., Goodman, A. A., Pineda, J. E., Halle, M. W. 2010, ApJ, 715, 1170
Arce, H. G., Shepherd, D., Gueth, F., Lee, C.-F., Bachiller, R., Rosen, A., Beuther, H. 2007, in Reipurth, B., Jewitt, D., Keil, K., eds, Protostars and Planets V, University of Arizona, p245
Bally, J., Devine, D., Reipurth, B. 1996, ApJ, 473, 49
Bally, J., Reipurth, B., Davis, C. J. 2007, in Reipurth, B., Jewitt, D., Keil, K., eds, Protostars and Planets V, University of Arizona, p215
Claussen, M. J., Marvel, K. B., Wootten, A., Wilking, B. A. 1998, ApJ, 507, L79
Claussen, M. J., Wilking, B. A., Benson, P. J., Wootten, A., Myers, P. C., Terebey, S. 1996, ApJS, 106, 111
Curtis, E. I., Richer, J. S., Swift, J. J., Williams, J. P. 2010, MNRAS, 408, 1516
Davis, C. J., Scholz, P., Lucas, P., Smith, M. D., Adamson, A. 2008, MNRAS, 387, 954
Fontani, F., Cesaroni, R., Furuya, R. S. 2010, A&A, 517, 56
Furuya, R. S., Kitamura, Y., Wootten, A., Claussen, M. J., Kawabe, R. 2003, ApJS, 144, 71
Goddi, C., Moscadelli, L. 2006, A&A, 447, 577
Gutermuth, R. A., Myers, P. C., Megeath, S. T., Allen, L. E., Pipher, J. L., Muzerolle, J., Porras, A., Winston, E., Fazio, G. 2008, ApJ, 674, 336
Hartmann, L., Ballesteros-Paredes, J., Bergin, E. A. 2001, ApJ, 562, 852

Haschick, A. D., Moran, J. M., Rodríguez, L. F., Burke, B. F., Greenfield, P., Garcia-Barreto, J. A. 1980, ApJ, 237, 26

Hatchell, J., Dunham, M. M. 2009, A&A, 502, 139

Hirota, T., Bushimata, T., Choi, Y. K., Honma, M., Imai, H., Iwadate, K., Jike, T., Kameya, O., Kamohara, R., Kan-Ya, Y., Kawaguchi, N., Kijima, M., Kobayashi, H., Kuji, S., Kurayama, T., Manabe, S., Miyaji, T., Nagayama, T., Nakagawa, A., Oh, C. S., Omodaka, T., Oyama, T., Sakai, S., Sasao, T., Sato, K., Shibata, K. M., Tamura, Y., Yamashita, K. 2008, PASJ, 60, 37

Honma, M., Bushimata, T., Choi, Y. K., Fujii, T., Hirota, T., Horiai, K., Imai, H., Inomata, N., Ishitsuka, J., Iwadate, K., Jike, T., Kameya, O., Kamohara, R., Kan-Ya, Y., Kawaguchi, N., Kijima, M., Kobayashi, H., Kuji, S., Kurayama, T., Manabe, S., Miyaji, T., Nakagawa, A., Nakashima, K., Oh, C. S., Omodaka, T., Oyama, T., Rioja, M., Sakai, S., Sato, K., Sasao, T., Shibata, K. M., Shimizu, R., Sora, K., Suda, H., Tamura, Y., Yamashita, K. 2005, PASJ, 57, 595
Ho, P. T. P., Barrett, A. H. 1979, BAAS, 11, 401
Kalenskii, S. V., Johansson, L. E. B., Bergman, P., Kurtz, S., Hofner, P., Walmsley, C. M., Slysh, V. I. 2010, MNRAS, 405, 613
Kalenskii, S. V., Promyslov, V. G., Slysh, V. I., Bergman, P., Winnberg, A. 2006, ARep, 50, 289
Knee, L. B. G., Sandell, G. 2000, A&A, 361, 671
Kurtz, S., Hofner, P., Álvarez, C. V. 2004, ApJS, 155, 149 Lada, C. J., Alves, J., Lada, E. A. 1996, AJ, 111, 1964

Looney, Leslie W., Mundy, Lee G., Welch, W. J. 2000, ApJ, 529, 477

Marvel, Kevin B., Wilking, Bruce A., Claussen, Mark J., Wootten, Alwyn 2008, ApJ, 685, 285
Matzner, C. D., McKee, C. F. 2000, ApJ, 545, 364
Moscadelli, L., Cesaroni, R., Rioja, M. J. 2005, A&A, 438, 889

Moscadelli, L., Testi, L., Furuya, R. S., Goddi, C., Claussen, M., Kitamura, Y., Wootten, A. 2006, A&A, 446, 985

Plunkett, A. L., Arce, H. G., Corder, S. A., Mardones, D., Sargent, A. I., Schnee, S. L. 2013, ApJ, 774, 22

Rebull, L. M., Stapelfeldt, K. R., Evans, N. J., II, Jørgensen, J. K., Harvey, P. M., Brooke, T. Y., Bourke, T. L., Padgett, D. L., Chapman, N. L., Lai, S.-P., Spiesman, W. J., Noriega-Crespo, A., Mern, B., Huard, T., Allen, L. E., Blake, G. A., Jarrett, T., Koerner, D. W., Mundy, L. G., Myers, P. C., Sargent, A. I., van Dishoeck, E. F., Wahhaj, Z., Young, K. E. 2007, ApJSS, 171, 447

Rodríguez, L. F., Anglada, G., Curiel, S. 1997, ApJ, 480, 125

Sakai, N., Ceccarelli, C., Bottinelli, S., Sakai, T., Yamamoto, S. 2012, ApJ, 754, 70

Sandell, G., Aspin, C., Duncan, W. D., Russell, A. P. G., Robson, E. I. 1991, ApJ, 376, 17

Sandell, G., Knee, L. B. G. 2001, ApJ, 546, 49

Sanna, A., Reid, M. J., Carrasco-González, C., Menten, K. M., Brunthaler, A., Moscadelli, L., Rygl, K. L. J. 2012, ApJ, 745, 191

Stutzki, J., Güsten, R. 1990, ApJ, 356, 513

Torrelles, J. M., Patel, N. A., Curiel, S., Estalella, R., Gómez, J. F., Rodríguez, L. F., Cantó, J., Anglada, G., Vlemmings, W., Garay, G., Raga, A. C., Ho, P. T. P. 2011, MNRAS, 410, 627

Torrelles, J. M., Patel, N. A., Gómez, J. F., Ho, P. T. P., Rodríguez, L. F., Anglada, G., Garay, G., Greenhill, L., Curiel, S., Cantó, J. 2001a, Nature, 411, 277

Torrelles, J. M., Patel, N. A., Gómez, J. F., Ho, P. T. P., Rodríguez, L. F., Anglada, G., Garay, G., Greenhill, L., Curiel, S., Cantó, J. 2001b, ApJ, 560, 853

Torrelles, J. M., Trinidad, M. A., Curiel, S., Estalella, R., Patel, N. A., Gómez, J. F., Anglada, G., Carrasco-González, C., Cantó, J., Raga, A., Rodríguez, L. F. 2014, MNRAS, 437, 3803

Trinidad, M. A., Curiel, S., Estalella, R., Cantó, J., Raga, A., Torrelles, J. M., Patel, N. A., Gómez, J. F., Anglada, G., Carrasco-González, C., Rodrguez, L. F. 2013, MNRAS, 430, 1309

Wootten, A., Claussen, M., Marvel, K., Wilking, B. 2002, IAU Symposium, 206, p100